\documentclass[leqno]{article} 
\usepackage{amssymb} 

\newcommand{\beq}{\begin{eqnarray}}
\newcommand{\eeq}{\end{eqnarray}}
\newcommand{\beqnn}{\begin{eqnarray*}}
\newcommand{\eeqnn}{\end{eqnarray*}}
\newtheorem{theorem}{Theorem}
\newtheorem{lemma}{Lemma}

\newcommand{\proof}{\paragraph*{{\it Proof.\/}}} 

\newcommand{\qed}{\fbox{\phantom{-}}\bigskip} 
\newcommand{\rd}{\partial}
\renewcommand{\Im}{\mathop{\mathrm{Im}}\nolimits}

\newcommand{\Tr}{\mathop{\mathrm{Tr}}}

\newcommand{\tp}[1]{\:{}^{\mathrm{t}}#1}
\newcommand{\Ker}{\mathop{\mathrm{Ker}}}
\newcommand{\Coker}{\mathop{\mathrm{Coker}}}

\newcommand{\CC}{\mathbf{C}}
\newcommand{\PP}{\mathbf{P}}

\newcommand{\ZZ}{\mathbf{Z}}
\newcommand{\fkg}{\mathfrak{g}} 
\newcommand{\bfalpha}{\mbox{\boldmath$\alpha$}}
\newcommand{\bfbeta}{\mbox{\boldmath$\beta$}}
\begin{document}



\title{Elliptic spectral parameter and \\
infinite dimensional Grassmann variety
\footnote{Contribution to the proceedings of the workshop 
``Infinite dimensional algebras and quantum integrable systems,''
Faro, Portugal, July 21-25, 2003.}}
\author{Kanehisa Takasaki\\
\normalsize 
Graduate School of Human and Environmental Studies, 
Kyoto University\\
\normalsize Yoshida, Sakyo, Kyoto 606-8501, Japan\\
\normalsize takasaki@math.h.kyoto-u.ac.jp}
\date{}
\maketitle

\begin{abstract}
Recent results on the Grassmannian perspective of soliton 
equations with an elliptic spectral parameter are presented 
along with a detailed review of the classical case with 
a rational spectral parameter.  The nonlinear Schr\"odinger 
hierarchy is picked out for illustration of the classical 
case.  This system is formulated as a dynamical system on 
a Lie group of Laurent series with factorization structure.  
The factorization structure induces a mapping to an infinite 
dimensional Grassmann variety.  The dynamical system on 
the Lie group is thereby mapped to a simple dynamical system 
on a subset of the Grassmann variety.  Upon suitable 
modification, almost the same procedure turns out to work 
for soliton equations with an elliptic spectral parameter.  
A clue is the geometry of holomorphic vector bundles over 
the elliptic curve hidden (or manifest) in the zero-curvature 
representation.  
\end{abstract}

\bigskip 
\begin{flushleft}
Keywords and phases: soliton equation, elliptic curve, 
holomorphic bundle, Grassmann variety\\
arXiv:nlin.SI/0312016
\end{flushleft}


\section{Introduction}

Since the first proposal two decades ago by Sato 
\cite{bib:SS82}, Segal and Wilson \cite{bib:SW85}, 
the Grassmannian perspective of soliton equations 
has been successful for a variety of cases, even including 
higher dimensional analogues such as the Bogomolny equation 
and the self-dual Yang-Mills equations \cite{bib:Ta84}.  
The fundamental observation of this perspective is that 
a soliton equation can be translated to a simple (essentially 
linear) dynamical system on a subset of an infinite dimensional 
``universal'' Grassmann variety.  Almost all of the cases 
thus examined, however, are equations with a {\it rational} 
zero-curvature representation, namely, equations whose 
zero-curvature equation is made of matrices depending 
rationally on a spectral parameter.  The status of 
soliton equations related to an elliptic or higher genus 
algebraic curve has still remained rather obscure, though 
a few notable studies \cite{bib:DJKM83,bib:CHMS93} were done 
on the Landau-Lifshitz equation (a typical soliton equation 
with a zero-curvature representation made of elliptic functions 
\cite{bib:Sk79,bib:Ch81}). 

Recent advances \cite{bib:BF01,bib:Kr02,bib:LOZ03} have revealed 
the existence of a wide class of new integrable PDE's with 
a zero-curvature representation constructed on an algebraic 
curve of arbitrary genus.  These equations, too, may be called 
``soliton equations'' in a loose sense, namely, without implying 
the existence of soliton or soliton-like solutions.  
The works of Ben-Zvi and Frenkel \cite{bib:BF01} and 
Levin, Olshanetsky and Zotov \cite{bib:LOZ03} both stem from 
the notion of the Hitchin systems \cite{bib:Hi90}, and aim to 
obtain an integrable PDE as a $1 + 1$ dimensional analogue of 
the Hitchin systems.  On the other hand, Krichever \cite{bib:Kr02} 
uses the so called Tyurin parameters to construct Lax or 
zero-curvature equations on an algebraic curve.  The notion 
of Tyurin parameters originates in algebraic geometry of 
holomorphic vector bundles over algebraic curves \cite{bib:Ty67}, 
and was applied by Krichever and Novikov in 1970's to 
the study of commutative rings of differential operators 
\cite{bib:Kr78,bib:KN78,bib:KN80}.  Krichever and Levin et al. 
illustrate their general scheme with several examples 
related to an elliptic curve.  These examples can be used as 
valuable material for case studies.  

One will naturally ask whether these new ``soliton equations'' 
can be understood in the Grassmannian perspective.  
An affirmative answer to this question has been obtained 
in the simplest case \cite{bib:Ta03a,bib:Ta03b}, namely, 
a few examples that have a zero-curvature prepresentation 
with $2 \times 2$ matrices defined on an elliptic curve.  
Although this is indeed a case study, the upshot clearly shows 
that a similar result holds in a general and universal form.  
What distinguishes between the new and conventional soliton equations 
is the structure of a holomorphic bundle on the relevant 
algebraic curve.  The aforementioned new equations are 
accompanied by a nontrivial bundle, which plays a central role 
in both the zero-curvature representation and 
the Grassmannian perspective.  This article presents an outline 
of these results.  

This article is organized as follows.  The first half 
(Sections 2, 3 and 4) of this article is a review on conventional 
soliton equations with a rational zero-curvature representation.  
The nonlinear Schr\"odinger hierarchy is picked out for 
illustration.  This system consists of an infinite number of 
evolution equations including the nonlinear Schr\"odinger equation 
itself in the lowest $1 + 1$ dimensional sector.  
One can reformulate this system as a dynamical system on 
a Lie group of Laurent series with factorization structure.  
The factorization induces a mapping to an infinite dimensional 
Grassmann variety.  The dynamical system on the Lie group 
is thereby mapped to a simple dynamical system on a subset 
of the Grassmann variety.  This example shows a typical way 
the usual soliton equations are treated in the Grassmannian 
perspective.  The second half (Section 5, 6 and 7) 
of this article presents the results on elliptic analogues 
\cite{bib:Ta03a,bib:Ta03b}.  Two different types of 
elliptic analogues are considered here.  The first case 
is an elliptic analogue of the nonlinear Schr\"odinger hierarchy.  
This system is constructed along the line of Krichever's scheme 
based on Tyurin parameters.  The second case is concerned 
with the Landau-Lifshitz equation and an associated hierarchy 
of evolution equations.  In both cases, a variant of 
the factorization is formulated as a Riemann-Hilbert problem 
with respect to the holomorphic bundle structure, and used 
to define a mapping to an infinite dimensional Grassmann variety.

\section{Nonlinear Schr\"odinger hierarchy}

The construction of the nonlinear Schr\"odinger hierarchy 
starts from the first order matrix differential operator 
$\rd_x - A(\lambda)$,  where $A(\lambda)$ a $2 \times 2$ 
matrix of the form 
\beq
  A(\lambda) 
  = \left(\begin{array}{cc} 
      \lambda & u \\
      v & - \lambda 
    \end{array}\right). 
\eeq
$u$ and $v$ are fields on the one dimensional space,  
$u = u(x),\; v = v(x)$, and $\lambda$ is a rational 
spectral parameter.  From the point of view of affine 
Lie algebras, it is also natural to express $A(\lambda)$ as 
\beq
  A(\lambda) = J\lambda + A^{(1)}, 
\eeq
where 
\beqnn
  J = \left(\begin{array}{cc} 
      1 & 0 \\
      0 & -1 
      \end{array}\right), \quad 
  A^{(1)} = \left(\begin{array}{cc} 
            0 & u \\
            v & 0 
            \end{array}\right). 
\eeqnn
Generalities and backgrounds of this kind of 
soliton equations can be found in Frenkel's lectures 
\cite{bib:Fr98}.

\subsection{Generating functions}

The first step of the formulation of the hierarchy is 
to construct a $2 \times 2$ matrix of generating functions 
\beqnn
  U(\lambda) = \sum_{n=0}^\infty U_n\lambda^{-n}, \quad 
  U_0 = J, 
\eeqnn
that satisfies the differential equation 
\beq
  [\rd_x - A(\lambda),\; U(\lambda)] = 0.  
\eeq
This reduces to the differential equations 
\beq
  \rd_xU_{n-1} = [J, U_n] + [A^{(1)}, U_{n-1}] 
\eeq
for $U_n$'s.  One can, in principle, solve these equations, 
by a subtle procedure decomposing the equations into the 
diagonal and off-diagonal part;  a similar procedure is 
used below to construct another generating function 
$\phi(\lambda)$.  This, however, leaves large arbitrariness 
in the solution.  Moreover, this is by no means an effective way.  

These problems are resolved by imposing 
the algebraic constraint 
\beq
  U(\lambda)^2 = I.  
\eeq
This amounts to finding $U(\lambda)$ in such a form as 
\beq
  U(\lambda) = \phi(\lambda)J\phi(\lambda)^{-1}, 
\eeq
where $\phi(\lambda)$ is another matrix of generating 
function 
\beqnn
  \phi(\lambda) = \sum_{n=0}^\infty \phi_n\lambda^{-n}, \quad 
  \phi_0 = I, 
\eeqnn
that satisfies the differential equation 
\beq
  \rd_x\phi(\lambda) 
  = A(\lambda)\phi(\lambda) - \phi(\lambda)J\lambda. 
\eeq
Remarkably, if the {\it existence} of a solution 
of this equation is ensured, one can uniquely determine 
$U_n$'s by a set of recurrence relations as follows. 

Note that the algebraic constraint implies 
the algebraic relations 
\beq
  0 = JU_n + U_nJ + \sum_{m=1}^{n-1} U_m U_{n-m}, 
  \quad n > 0, 
\eeq 
that hold for $U_n$'s.  Combining this with the differential 
equations 
\beqnn
  \rd_xU_{n-1} = JU_n - U_nJ + [A^{(1)},U_{n-1}], 
\eeqnn
one obtains the relations 
\beq
  2JU_n = \rd_xU_{n-1} - [A^{(1)},U_{n-1}] 
          - \sum_{m=1}^{n-1} U_m U_{n-m}. 
\eeq
These relations take the form of recurrence relations, 
which enables one to calculate $U_n$'s successively. 
The first few terms read 
\beqnn
  U_1 = \left(\begin{array}{cc}
        0 & u \\
        v & 0 
        \end{array}\right), 
  \quad 
  U_2 = \left(\begin{array}{ll} 
        - \frac{1}{2}uv & \frac{1}{2}\rd_xu \\
        - \frac{1}{2}\rd_xv & \frac{1}{2}uv 
        \end{array}\right), 
  \quad 
  \mbox{etc.}
\eeqnn
The matrix elements of $U_n$ thus obtained 
are ``local'' quantities, namely, polynomials 
of derivatives of $u,v$.  

What is left is to prove that the second generating function  
$\phi(\lambda)$ does exist.  The equations for the Laurent 
coefficients of $\phi(\lambda)$ read 
\beq
  \rd_x \phi_n = [J, \phi_{n+1}] + A^{(1)}\phi_n. 
\eeq
One can split this matrix equation into the diagonal 
and off-diagonal part.  This results in the two equations 
\beqnn
  \rd_x(\phi_n)_{\mathrm{diag}} 
  = (A^{(1)}\phi_n)_{\mathrm{diag}} 
\eeqnn
and 
\beqnn
  \rd_x(\phi_n)_{\mathrm{off-diag}} 
  = [J,\; (\phi_{n+1})_{\mathrm{off-diag}}] 
    + (A^{(1)}\phi_n)_{\mathrm{off-diag}} 
\eeqnn
for the diagonal part $(\phi_n)_{\mathrm{diag}}$ 
and the off-diagonal parts $(\phi_n)_{\mathrm{off-diag}}$ 
of $\phi_n$.  The first equation determines the diagonal 
part of $\phi_n$ up to integration constants. 
The second equation is rather an algebraic equation 
that determines the off-diagonal part of $\phi_{n+1}$ 
from the lower coefficients $\phi_1,\cdots,\phi_n$.  
One can thus construct a solution of these equations.  

Note that, unlike the aforementioned construction 
of $U_n$'s, the construction of $\phi(\lambda)$ 
is not purely algebraic (the outcome is accordingly 
``nonlocal'') and leaves large arbitrariness.  It is, 
however, $\phi(\lambda)$ rather than $U(\lambda)$ 
that plays a more fundamental role in the passage 
to the Grassmannian perspective.

\subsection{Formulation of hierarchy}

Let $t = (t_1,t_2,\ldots)$ be a sequence of ``time'' variables; 
the first one $t_1$ is to be identified with the spatial 
variable $x$.  The $n$-th time evolution is generated by 
the matrix 
\beq
  A_n(\lambda) 
  = \sum_{m=0}^n U_m \lambda^{n-m} 
  = \bigl(U(\lambda)\lambda^n\bigr)_{+}. 
\eeq
Here $(\cdot)_{+}$ denotes the polynomial part of 
a Laurent series of $\lambda$.  
Having introduced these matrices, one can define 
the nonlinear Schr\"odinger hierarchy as 
the system of the Lax equations 
\beq
  [\rd_{t_n} - A_n(\lambda),\; U(\lambda)] = 0 \quad 
\eeq
for $n = 1,2,\ldots$.  Since $A_1(\lambda) = A(\lambda)$, 
one can identify $t_1$ with $x$.  In many aspects, 
this formulation of the nonlinear Schr\"odinger hierarchy 
resembles the formulation of the KP hierarchy \cite{bib:SS82}.  
For instance, as known in the case of the KP hierarchy, 
the system of Lax equations is equivalent to the system of 
zero-curvature equations 
\beq
  [\rd_{t_m} - A_m(\lambda),\; \rd_{t_n} - A_n(\lambda)] = 0  
\eeq
for $m,n = 1,2,\ldots$, namely, one can derive one 
from the other. The zero-curvature equation 
for $m= 1$ and $n = 2$ gives the equations 
\beqnn
  \rd_t u - \frac{1}{2}\rd_x^2 u + u^2v = 0, \quad 
  \rd_t v + \frac{1}{2}\rd_x^2 v - uv^2 = 0, 
\eeqnn
which turn into the usual nonlinear Schr\"odinger 
equation by rescaling the variables as 
$u \to e^{at}u$, $v \to e^{-at}v$, $t \to it$ 
($a = \pm 1$) and imposing the reality condition 
$v = \overline{u}$. 

Yet another formulation of the hierarchy is achieved 
by the system of differential equations 
\beq
  \rd_{t_n}\phi(\lambda) 
  = A_n(\lambda)\phi(\lambda) - \phi(\lambda)J\lambda^n. 
  \label{eq:rd-tn-phi1}
\eeq 
If one introduces the co called ``formal Baker-Akhiezer function'' 
\beq
  \psi(\lambda) 
  = \phi(\lambda)\exp\Bigl(\sum_{n=1}^\infty t_nJ\lambda^n\Bigl), 
\eeq
the foregoing equations turn into the auxiliary linear equations 
\beq
  \rd_{t_n}\psi(\lambda) = A_n(\lambda)\psi(\lambda). 
\eeq
The Frobenius integrability condition of these equations 
yields the aforementioned zero-curvature equations.  
On the other hand, if one rewrites the definition of 
$A_n(\lambda)$ as 
\beq
  A_n(\lambda) 
  = \Bigl(\phi(\lambda)J\lambda^n\phi(\lambda)^{-1}\Bigr)_{+} 
\eeq
and insert it into the differential equations for 
$\phi(\lambda)$, the outcome is a system of nonlinear 
evolution equations for $\phi(\lambda)$ of the form 
\beq
  \rd_{t_n}\phi(\lambda) 
  = - \Bigl(\phi(\lambda)J\lambda^n\phi(\lambda)^{-1}\Bigr)_{-} 
      \phi(\lambda), 
  \label{eq:rd-tn-phi2} 
\eeq
where $(\cdot)_{-}$ denotes the negative power part 
of a Laurent series of $\lambda$.  These equations 
may be thought of as the most fundamental 
because the Lax and zero-curvature equations can be 
derived from these equations.

\section{Nonlinear Schr\"odinger hierarchy as dynamical system 
on Lie group of Laurent series} 

The nonlinear Schr\"odinger hierarchy can be interpretend 
as a dynamical system on an infinite dimensional Lie group.  
It is customary to formulate such a statement in terms of 
a loop group, namely, the set of a suitable class of 
(smooth, real-analytic or square-integrable) mappings from 
$S^1 = \{ \lambda \in \CC \mid |\lambda^{-1}| = a \}$ 
to $\mathrm{SL}(2,\CC)$.   From an aesthetic point of view, 
however, fixing a circle is not beautiful; the circle is 
a kind of artifact that did not exist in the formulation 
of the hierarchy itself.  A better approach is to use 
a Lie group of Laurent series that converge in a neighborhood 
of $\lambda = \infty$ except at the point $\lambda = \infty$.

\subsection{Lie algebras and groups of Laurent series}

Let $\fkg$ denote the Lie algebra of Laurent series 
of the form 
\beq
  X(\lambda) = \sum_{n=-\infty}^\infty X_n\lambda^n, \quad 
  X_n \in \mathrm{sl}(2,\CC), 
\eeq
that converge in a neighborhood of $\lambda = \infty $ 
except at $\lambda = \infty$.  In other words, 
the coefficients are assumed to satisfy the conditions 
\beqnn
  \lim_{n\to\infty}|X_n|^{1/n} = 0, \quad 
  \limsup_{n\to\infty}|X_{-n}|^{1/n} < \infty. 
\eeqnn
This Lie algebra has the direct sum decomposition 
\beq
  \fkg = \fkg_{+} \oplus \fkg_{-}, 
\eeq
where $\fkg_\pm$ are subalgebras of the form 
\beqnn
  \fkg_{+} = \{ X(\lambda) \in\fkg \mid 
    \mbox{$X_n = 0$ for $n < 0$}\}, 
  \\
  \fkg_{-} = \{ X(\lambda) \in\fkg \mid 
    \mbox{$X_n = 0$ for $n \ge 0$}\}. 
\eeqnn

The direct sum decomposition induces a factorization of 
the associated Lie group $G = \exp\fkg$ to the subgroups 
$G_\pm = \exp\fkg_\pm$, namely, any element $g(\lambda)$ 
of $G$ near the unit matrix $I$ can be uniquely 
factorized as 
\beq
  g(\lambda) = g_{+}(\lambda)^{-1}g_{-}(\lambda), 
  \quad 
  g_\pm(\lambda) \in G_\pm. 
\eeq
Analytically, this is nothing but the so called 
Riemann-Hilbert problem.  In geometric terms, $g(\lambda)$ 
is the transition function of a holomorphic $\mathrm{SL}(2,\CC)$ 
bundle $P$ over $\PP^1$ obtained by gluing trivial bundles over 
two disks $D_{+}$ and $D_{-}$, $D_{+} \cup D_{-} = \PP^1$, as 
\beq
  P = D_{+} \times \mathrm{SL}(2,\CC) \sqcup 
      D_{-} \times \mathrm{SL}(2,\CC) \;/\; \sim, 
\eeq
where $(\lambda,g_{+}) \in D_{+}\times\mathrm{SL}(2,\CC)$ 
and $(\lambda,g_{-}) \in D_{-}\times\mathrm{SL}(2,\CC)$ are
identified if $g_{+} = g(\lambda)g_{-}$.  Factorizability 
of $g(\lambda)$ amounts to holomorphic triviality of $P$ and 
of associated vector bundles.

\subsection{Factorization method}

Let $\phi(\lambda)$ denote an arbitrary element of $G_{-}$ 
and consider the factorization problem 
\beq
  \phi(\lambda)\exp\Bigl(- \sum_{n=1}^\infty t_nJ\lambda^n\Bigr) 
  = \chi(t,\lambda)^{-1}\phi(t,\lambda), 
  \label{eq:factor-chi-phi}\\ 
  \chi(t,\lambda) \in G_{+}, \quad 
  \phi(t,\lambda) \in G_{-}. \nonumber 
\eeq

\begin{lemma}
If $t$ is sufficiently small, the factorization problem 
(\ref{eq:factor-chi-phi}) has a unique solution.  
\end{lemma}

\proof 
Since $\phi(t,\lambda)$ is expected to be 
a small deformation of $\phi(\lambda)$, one can 
assume it in the form 
\beqnn
  \phi(t,\lambda) = \tilde{\chi}(t,\lambda)\phi(\lambda), \quad 
  \tilde{\chi}(t,\lambda) \in G_{-}, 
\eeqnn
and convert the problem to the form 
\beqnn
  \phi(\lambda)\exp\Bigl(- \sum_{n=1}^\infty t_nJ\lambda^n\Bigr) 
  \phi(\lambda)^{-1} 
  = \chi(t,\lambda)^{-1}\tilde{\chi}(t,\lambda). 
\eeqnn
If $t$ is sufficiently small, the left hand side 
is close to the unit matrix $I$, so that 
one can resort to the local factorizability of $G$.  
\qed 

Suppose that the factorization problem (\ref{eq:factor-chi-phi}) 
does have a unique solution.  The second factor $\phi(t,\lambda)$ 
then turns out to give a solution of (\ref{eq:rd-tn-phi2}): 

\begin{theorem}
The second factor $\phi(t,\lambda)$ of the factorization problem 
satisfies the evolution equations (\ref{eq:rd-tn-phi2}) and 
the initial condition $\phi(0,\lambda) = \phi(\lambda)$. 
\end{theorem}

\proof
If one rewrites the factorization relation as 
\beqnn
  \chi(t,\lambda)\phi(\lambda) 
  = \phi(t,\lambda)\exp\Bigl(\sum_{n=1}^\infty t_nJ\lambda^n\Bigr) 
\eeqnn
and differentiate both hand sides by $t_n$, the outcome reads 
\beqnn
\lefteqn{\rd_{t_n}\chi(t,\lambda)\cdot\phi(\lambda)} 
  \nonumber \\
  &=& \rd_{t_n}\phi(t,\lambda)\cdot
      \exp\Bigl(\sum_{n=1}^\infty t_nJ\lambda^n\Bigr) 
    + \phi(t,\lambda)J\lambda^n
      \exp\Bigl(\sum_{n=1}^\infty t_nJ\lambda^n\Bigr). 
\eeqnn
One can use the previous relation once again 
to eliminate $\phi(\lambda)$ and the exponential 
from this relation.  This yields the relation 
\beqnn
  \rd_{t_n}\chi(t,\lambda)\cdot\chi(t,\lambda)^{-1} 
  = \rd_{t_n}\phi(t,\lambda)\cdot\phi(t,\lambda)^{-1}
    + \phi(t,\lambda)J\lambda^n\phi(t,\lambda)^{-1}.  
\eeqnn
Let $A_n(t,\lambda)$ denote the $2 \times 2$ matrix defined 
by both hand sides of the last equation.  This leads to 
the two expressions 
\beqnn
  A_n(t,\lambda) 
  = \rd_{t_n}\chi(t,\lambda)\cdot\chi(t,\lambda)^{-1}
\eeqnn
and 
\beqnn
  A_n(t,\lambda) 
  = \rd_{t_n}\phi(t,\lambda)\cdot\phi(t,\lambda)^{-1} 
    + \phi(t,\lambda)J\lambda^n\phi(t,\lambda)^{-1} 
\eeqnn
of $A_n(t,\lambda)$.  The first expression shows 
that $A_n(t,\lambda)$ takes values in $\fkg_{+}$, 
so that one can replace the right hand side of 
the second expression by its projection 
onto $\fkg_{+}$.  Since the first term 
$\rd_{t_n}\phi(t,\lambda)\cdot\phi(t,\lambda)^{-1}$ 
obviously disappears upon projection, one finds that 
\beqnn
  A_n(\lambda) 
  = \Bigl(\phi(t,\lambda)J\lambda^n\phi(t,\lambda)^{-1}
    \Bigr)_{+}. 
\eeqnn
These results show that $\phi(t,\lambda)$ does satisfy 
(\ref{eq:rd-tn-phi2}) as expected.  Uniqueness of 
the factorization implies that $\phi(0,\lambda) = \phi(\lambda)$. 
\qed

This result can be restated in geometric terms as follows.   
Evolution equations (\ref{eq:rd-tn-phi2}) define 
a dynamical system on $G_{-}$.  Factorizability of $G$ 
implies that 
\beq
\begin{array}{ccc}
  G_{-} &\to& G_{+}\backslash G \\
  \phi(\lambda) &\mapsto& G_{+}\phi(\lambda) 
  \label{eq:Gminus-into-coset}
\end{array}
\eeq
is an injective mapping with open (and dense) image. 
The dynamical system on $G_{-}$ is nothing but 
the pullback, by (\ref{eq:Gminus-into-coset}), 
of the exponential flows 
\beq
  G_{-}g(\lambda) \mapsto 
  G_{-}g(\lambda)\exp\Bigl(- \sum_{n=1}^\infty t_nJ\lambda^n\Bigr) 
\eeq
on the coset $G_{+}\backslash G$.  Moreover, 
the coset $G_{+}\backslash G$ may be interpreted as 
the moduli space of holomorphic $\mathrm{SL}(2,\CC)$ bundles 
over $\PP^1$ equipped with trivialization over $D_{-}$, 
elements of the form $G_{+}\phi(\lambda)$ being a representative 
of trivial bundles.

\section{Nonlinear Schr\"odinger hierarchy as dynamical system 
on infinite dimensional Grassmann variety}

The forgoing dynamical system on $G_{-}$ can be mapped 
to a dynamical system in an infinite dimensional 
Grassmann variety.  In the literature, two different models 
of Grassmann varieties (or Grassmann manifolds) 
have been used for this kind of description.  
One is Sato's algebraic or complex analytic model based on 
a vector space of Laurent series \cite{bib:SS82}; 
the other is Segal and Wilson's functional analytic model 
made from the Hilbert space of square-integrable functions 
on a circle \cite{bib:SW85}.  One should obviously choose 
Sato's model in the present setting.

\subsection{Formulation of Grassmann variety}

The Grassmann variety $\mathrm{Gr}$ to be used below 
is constructed from the vector space $V$ of $2 \times 2$ 
matrices $X(\lambda)$ of Laurent series of the form 
\beq
  X(\lambda) = \sum_{n=-\infty}^\infty X_n\lambda^n, \quad 
  X_n \in \mathrm{gl}(2,\CC), 
\eeq
that converge in a neighborhood of $\lambda = \infty$ 
except at the point $\lambda = \infty$.  This is almost 
the same thing as $\fkg$ but the coefficients $X_n$ 
are now an arbitrary matrix; recall that $\mathrm{gl}(2,\CC)$ 
denotes the vector space (or matrix Lie algebra) of 
arbitrary $2 \times 2$ complex matrices.   
This vector space is a matrix version of the vector space 
$V^{\mathrm{ana}(\infty)}$ in Sato's list of models 
\cite{bib:SS82}.  As noted therein, this vector space 
has a natural linear topology.  The Grassmann variety 
$\mathrm{Gr}$ consists of closed vector subspaces 
$W \subset V$ satisfying an additional condition 
as follows: 
\beq
\lefteqn{\mathrm{Gr} = \{ W \subset V \mid} \\
  &&  \dim\Ker(W \to V/V_{-}) 
      = \dim\Coker(W \to V/V_{-}) < \infty \}. 
  \nonumber
\eeq
Here $V_{-}$ denotes the vector subspace of $V$ 
consisting of $X(\lambda)$'s that contain only 
negative powers of $\lambda$: 
\beq
  V_{-} = \{ X(\lambda) \in V \mid 
            \mbox{$X_n = 0$ for $n \ge 0$}\}. 
\eeq
The map $W \to V/V_{-}$ is the composition of 
the inclusion $W \hookrightarrow V$ and 
the canonical projection $V \to V/V_{-}$.  
The so called ``big cell'' of $\mathrm{Gr}$ 
consists of subspaces $W \subset V$ for which 
this linear map is an isomorphism: 
\beq
  \mathrm{Gr}^\circ 
  = \{ W \in \mathrm{Gr} \mid W \simeq V/V_{-}\}. 
\eeq
This is an open subset of $\mathrm{Gr}$, namely, 
sufficiently small deformations of any element of 
$\mathrm{Gr}^\circ$ remains in $\mathrm{Gr}^\circ$.

\subsection{Vacuum and dressing} 
 
Let $W_0$ be the subspace of $V$ spanned by nonnegative 
powers of $\lambda$: 
\beq
  W_0 = \{ X(\lambda) \in V \mid 
            \mbox{$X_n = 0$ for $n < 0$}\}. 
\eeq
The linear map $W_0 \to V/V_{-}$ is obviously isomorphic
in view of the basis $\{E_{ij}\lambda^n \mid n \ge 0,\; 
i,j = 1,2\}$ for both vector spaces ($E_{ij}$ are 
the standard basis of $\mathrm{gl}(2,\CC)$).  Hence 
$W_0$ is an element of the big cell $\mathrm{Gr}^\circ$. 
This special element of the big cell plays the role of 
``vacuum,''  which corresponds to the vacuum solution 
$u = v = 0$ of the nonlinear Schr\"odinger equation.  

One can ``dress'' $W_0$ by an arbitrary element of $G_{-}$: 
\beq
  W = W_0\phi(\lambda), \quad 
  \phi(\lambda) = I + \sum_{n=1}^\infty \phi_n\lambda^{-n} 
  \in G_{-}. 
\eeq

\begin{lemma} \label{lem:W-in-bigcell}
$W$ is an element of the big cell $\mathrm{Gr}^\circ$.  
\end{lemma}

\proof 
$W$ is spanned by $E_{ij}\lambda^n\phi(\lambda)$, 
$n \ge 0$, $i,j = 1,2$.   By a triangular linear 
transformation, one can modify this basis of $W$ 
to another basis $\{w_{n,ij}(\lambda) \mid 
n \ge 0,\; i,j = 1,2\}$ such that 
\beqnn
  w_{n,ij}(\lambda) = E_{ij}\lambda^n + O(\lambda^{-1}). 
\eeqnn
More explicitly, 
\beqnn
  w_{n,ij}(\lambda) 
  = \Bigl(\phi(\lambda)E_{ij}\lambda^n\phi(\lambda)^{-1}
    \Bigr)_{+} \phi(\lambda). 
\eeqnn
The linear map $W \to V/V_{-}$ sends this basis to 
the standard basis $\{E_{ij}\lambda^n \mid n \ge 0,\; 
i,j = 1,2\}$ of $V/V_{-}$, thereby turns out to be 
an isomorphism. 
\qed 

The phase space $G_{-}$ of the dynamical system 
of the last section can be thus mapped, 
by the correspondence 
\beq
  \phi(\lambda) \mapsto W = W_0 \phi(\lambda), 
\eeq
to the set 
\beq
  \mathcal{M} 
  = \{ W \in \mathrm{Gr}^\circ \mid 
       W = W_0\phi(\lambda), \; \phi(\lambda) \in G_{-} \} 
\eeq
of these ``dressed vacua'' in (the big cell of) 
the infinite dimensional Grassmann variety $\mathrm{Gr}$.  
The problem to be addressed next is to describe 
the dynamical motion on this new phase space.  

Actually, the foregoing mapping $G_{-} \stackrel{\sim}{\to} 
\mathcal{M}$ can be understood in a slightly more general form.  
Namely, the mapping can be extended to 
\beq
\begin{array}{ccc}
  G_{+}\backslash G &\to& \mathrm{Gr} \\
  G_{+}g(\lambda) &\mapsto& W_0g(\lambda) 
\end{array}
\eeq
that sends the coset $G_{+}\backslash G$ into 
$\mathrm{Gr}$.  Note that this mapping is well defined 
and injective because 
\beqnn
  g(\lambda) \in G_{+} \;\Longleftrightarrow\; 
  W_0g(\lambda) = W_0 
\eeqnn
(cf. Lemma \ref{lem:W0-chi}).  Thus, combined with 
the open embedding (\ref{eq:Gminus-into-coset}) 
of $G_{-}$ into $G_{+}\backslash G$,  
the mapping $G_{-} \stackrel{\sim}{\to} \mathcal{M}$ 
is substantially the well known embedding of 
the ``affine Grassmannian'' $G_{+}\backslash G$ into 
the Sato Grassmannian \cite{bib:SW85}.

\subsection{Dynamical system on space of dressed vacua} 

For simplicity, the following consideration is limited 
to small values of $t$.  The factorization problem 
(\ref{eq:factor-chi-phi}) is thereby ensured to have 
a unique solution.  The goal is to elucidate the motion 
of $W(t) = W_0\phi(t,\lambda) \in \mathcal{M}$.  
A clue to the answer is the following.  

\begin{lemma} \label{lem:W0-chi}
$W_0\chi(t,\lambda) = W_0$.  
\end{lemma}

\proof
$W_0$ is obviously closed under multiplication of two element. 
By construction, $\chi(t,\lambda)$ is obviously an element 
of $W_0$.  Therefore $W_0\chi(t,\lambda) \subseteq W_0$.  
On the other hand, the inverse $\chi(t,\lambda)^{-1}$ is 
also an element of $G_{+}$ as far as $t$ is sufficiently small, 
so that the same reasoning leads to the conclusion that 
$W_0\chi(t,\lambda) \subseteq W_0$.  Thus the equality follows. 
\qed 

If one rewrites the factorization relation 
(\ref{eq:factor-chi-phi}) as 
\beqnn
  \phi(t,\lambda) 
  = \chi(t,\lambda)\phi(0,\lambda)
    \exp\Bigl(- \sum_{n=1}^\infty t_nJ\lambda^n\Bigr) 
\eeqnn
and insert it into the definition 
$W(t) = W_0\phi(t,\lambda)$ of $W(t)$, one finds that 
\beqnn
  W(t) 
  &=& W_0\chi(t,\lambda)\phi(0,\lambda)
      \exp\Bigl(- \sum_{n=1}^\infty t_nJ\lambda^n\Bigr) 
  \nonumber \\
  &=& W_0\phi(0,\lambda)
      \exp\Bigl(- \sum_{n=1}^\infty t_nJ\lambda^n\Bigr) 
  \nonumber \\
  &=& W(0)\exp\Bigl(- \sum_{n=1}^\infty t_nJ\lambda^n\Bigr). 
\eeqnn
Note that the lemma has been used in the first stage; 
the first factor $\chi(t,z)$ of the factorization pair 
is absorbed by $W_0$.  Thus the motion of the point $W(t)$ 
of $\mathcal{M}$ turns out to obey the simple exponential law 
\beq
  W(t) = W(0)\exp\Bigl(- \sum_{n=1}^\infty t_nJ\lambda^n\Bigr). 
  \label{eq:exp-flow}
\eeq

One thus arrives at the following fundamental picture, 
which is an example of the Grassmannian perspective 
of soliton equations due to Sato \cite{bib:SS82} 
and Segal and Wilson \cite{bib:SW85}. 

\begin{theorem}
The nonlinear Schr\"odinger hierarchy can be mapped, 
by the correspondence $W(t) = W_0 \phi(t,\lambda)$, 
to a dynamical system on the set $\mathcal{M}$ of 
dressed vacua in the Grassmann variety $\mathrm{Gr}$.  
The motion of $W(t)$ obeys the exponential law 
(\ref{eq:exp-flow}).  
\end{theorem}

Conversely, given an arbitrary element $\phi(\lambda)$ of 
$G_{-}$, one can derive a solution of the factorization 
problem (\ref{eq:factor-chi-phi}) from this dynamical system. 
By Lemma \ref{lem:W-in-bigcell}, $W(0) = W_0\phi(\lambda)$ 
is an element of the big cell.  If $t$ is sufficiently small, 
the point $W(t)$ on the trajectory of the exponential flows 
(\ref{eq:exp-flow}) still remains in the big cell, 
because the big cell is an open subset of $\mathrm{Gr}$.  
This means that the linear map $W(t) \to V/V_{-}$, 
i.e., the composition of the inclusion 
$W(t) \hookrightarrow V$ and 
the canonical projection $V \to V/V_{-}$, 
is an isomorphism.  Let $\phi(t,\lambda) \in W(t)$ be 
the inverse image of $I \in V/V_{-}$ by this isomorphism. 
Being equal to $I$ modulo $V_{-}$, $\phi(t,\lambda)$ 
is a Laurent series of the form 
\beqnn
  \phi(t,\lambda) = I + \sum_{n=1}^\infty \phi_n(t)\lambda^{-n}, 
  \quad \phi_n(t) \in \mathrm{gl}(2,\CC). 
\eeqnn
On the other hand, as an element of 
\beqnn
  W(t) = W_0 \phi(\lambda) 
         \exp\Bigl(- \sum_{n=1}^\infty t_nJ\lambda^n\Bigr), 
\eeqnn
$\phi(t,\lambda)$ can also be expressed as 
\beqnn
  \phi(t,\lambda) 
  = \chi(t,\lambda)\phi(\lambda)
    \exp\Bigl(- \sum_{n=1}^\infty t_nJ\lambda^n\Bigr) 
\eeqnn
with an element $\chi(t,\lambda)$ of $W_0$.  
Taking the determinant of both hand sides of 
the last equality yields the equality 
\beqnn
  \det\phi(t,\lambda) = \det\chi(t,\lambda), 
\eeqnn
in which $\phi(\lambda)$ and the exponential disappear 
because they are known to be unimodular.  Notice here that 
\beqnn
  \det\phi(t,\lambda) 
  &=& 1 + (\mbox{negative powers of $\lambda$}), 
  \nonumber \\
  \det\chi(t,\lambda)
  &=& (\mbox{nonnegative powers of $\lambda$}). 
\eeqnn
Consequently, both hand sides of the determinant equality 
is actually equal to $1$.  This implies that 
$\phi(t,\lambda) \in G_{-}$ and $\chi(t,\lambda) \in G_{+}$, 
so that they give a solution of the factorization problem 
(\ref{eq:factor-chi-phi}).  

This shows another aspect of the Grassmannian perspective. 
Namely, the Grassmann variety can be used as a tool for 
solving a factorization or Rimann-Hilbert problem. 
This point of view turns out to be useful later.

\section{Elliptic analogue of nonlinear Schr\"odinger hierarchy}

We now turn to examples with an elliptic spectral parameter.  
The first example is based on an example of Krichever's general 
construction \cite{bib:Kr02}.  Let us briefly recall the background 
of Krichever's work. 

It is well known, after the work of Zakharov and Mikhailov 
\cite{bib:ZM82}, that a naive attempt at the construction 
of a zero-curvature equation 
\beqnn
  [\rd_x - A(P),\; \rd_t - B(P)] = 0, \quad P \in \Gamma, 
\eeqnn
on an arbitrary algebraic curve $\Gamma$ is confronted 
with a serious difficulty that stems from the Riemann-Roch theorem.  
If the construction for $\Gamma = \PP^1$ also works 
in the general case, $A(P)$ and $B(P)$ are a matrix of 
meromorphic functions on $\Gamma$ with fixed poles, 
say, $Q_1,\ldots,Q_s$ of order $m_1,\ldots,m_s$ for $A(P)$ 
and $n_1,\ldots,n_s$ for $B(P)$.  Choosing a suitable 
linearly independent set of meromorphic functions $f_j(P)$, 
$h = 1,\ldots,M$, and $g_k(P)$, $k = 1,\ldots,N$, one can expand 
$A(P)$ and $B(P)$ as 
\beqnn
  A(P) = \sum_{j=1}^M A_j f_j(P), \quad 
  B(P) = \sum_{k=1}^N B_k g_k(P). 
\eeqnn
The (matrix-valued) coefficients $A_j,B_k$ are interpreted 
as the field variables $A_j = A_j(x,t)$, $B_k = B_k(x,t)$, 
for which the zero-curvature equation induces a set of PDE's.  
Part of these field variables can be eliminated by gauge 
transformations $A_j \to g^{-1}A_jg - g_xg^{-1}$, 
$B_k \to g^{-1}B_kg - g_tg^{-1}$.  In the case where $\Gamma = \PP^1$, 
suitable gauge fixing leads to a determined system of PDE's (i.e., 
a system of evolution equations) for the reduced field variables.  
In contrast, if the genus of $\Gamma$ is not zero, 
the Riemann-Roch theorem implies that the zero-curvature equation 
in a ``general position'' is an overdetermined system for $A_j$'s 
and $B_k$'s.  This means that one has to assume some special structure 
in $A(P)$ and $B(P)$ to obtain a consistent system of evolution 
equations.  An example is the Landau-Lifshitz equation (for which 
$\Gamma$ is an elliptic curve).  

Krichever \cite{bib:Kr02} pointed out that this difficulty 
can be avoided by allowing $A(P),B(P)$ to have extra 
``movable'' poles $\gamma_s$ at which the solutions of 
the auxiliary linear system $\rd_x\psi(P) = A(P)\psi(P),\; 
\rd_t\psi(P) = B(P)\psi(P)$ remain regular.  
This is reminiscent of the notion of ``apparent singularities'' 
in the theory of ordinary differential equations.  
The number of necessary movable poles turns out to be equal 
to $rg$, where $r$ is the size of the matrices $A(P),B(P)$ 
and $g$ the genus of $\Gamma$.  Moreover, to each movable pole 
is assigned a directional vector $\bfalpha_s \in \PP^{r-1}$ 
as an extra parameter.  These pairs $(\gamma_s,\bfalpha_s)$, 
$s =1,\ldots,rg$, are called ``Tyurin parameters'' and 
now join the game as new dynamical variables.  
The elliptic analogue of the nonlinear Schr\"odinger hierarchy 
amounts to the case where $g = 1$ and $r = 2$.

\subsection{Matrix of elliptic functions parametrized by 
Tyurin parameters}

The first stage of construction is to choose a suitable 
counterpart $A(z)$ of $A(\lambda)$.  This is a $2 \times 2$ 
matrix of elliptic functions on a nonsingular elliptic curve. 
The spectral parameter $z$ is now understood to be 
the standard complex coordinate on the torus $\Gamma 
= \CC/(2\omega_1\ZZ + 2\omega_3\ZZ)$ that realizes 
the elliptic curve.  

In addition to a pole at $z = 0$ (which corresponds to 
$\lambda = \infty$ in the case of the nonlinear Schr\"odinger 
hierarchy), this matrix has two extra poles 
$\gamma_1,\gamma_2$, $\gamma_1 \not= \gamma_2$, 
that depend on $x$ and $t_n$'s.  
Two directional vectors $\bfalpha_1,\bfalpha_2 \in \PP^1$ 
are introduced as the other half of the Tyurin parameters.  
These directional vectors can be normalized as $\bfalpha_s 
= \tp{(\alpha_s,1)}$.  The two fields $u,v$ in the usual 
nonlinear Schr\"odinger hierarchy also appear here.  
Thus one has altogether six dynamical variables $u,v,
\gamma_1,\gamma_2,\alpha_1,\alpha_2$ in the formulation 
of this elliptic analogue.  

The matrix $A(z)$ is defined, indirectly, by the following 
properties: 
\begin{enumerate}
\item $A(z)$ has poles at $z = 0,\gamma_1,\gamma_2$ and 
is holomorphic at other points.  
\item As $z \to 0$, 
\beqnn
   A(z) = \left(\begin{array}{cc} 
          z^{-1} & u \\
          v & - z^{-1} 
          \end{array}\right) 
        + O(z). 
\eeqnn
\item As $z \to \gamma_s$, $s = 1,2$, 
\beqnn
  A(z) = \frac{\bfbeta_s\tp{\bfalpha_s}}{z - \gamma_s} + O(1), 
\eeqnn
where $\bfalpha_s$ and $\bfbeta_s$ are two-dimensional 
column vectors that do not depend on $z$.  $\bfalpha_s$ 
is normalized as $\bfalpha_s = \tp{(\alpha_s,1)}$.  
\end{enumerate}

\begin{lemma}
If $\alpha_1 \not= \alpha_2$, a matrix $A(z)$ of meromorphic 
functions on $\Gamma$ with these properties does exists.  
It is unique and can be written explicitly 
in terms of the Weierstrass zeta function $\zeta(z)$ as 
\beq
  A(z) 
  = \sum_{s=1,2}\bfbeta_s\tp{\bfalpha_s}
      (\zeta(z - \gamma_s) + \zeta(\gamma_s)) 
  + \left(\begin{array}{cc} 
      \zeta(z) & u \\
      v & - \zeta(z) 
    \end{array}\right), 
\eeq
where 
\beqnn
  \bfbeta_1 
  = \frac{1}{\alpha_1 - \alpha_2}
    \left(\begin{array}{c}
      - 1 \\
      - \alpha_2 
    \end{array}\right), \quad 
  \bfbeta_2 
  = \frac{1}{\alpha_1 - \alpha_2} 
    \left(\begin{array}{c}
      1 \\
      \alpha_1 
    \end{array}\right).      
\eeqnn
\end{lemma}

\proof 
One can express $A(z)$ as 
\beqnn
  A(z) = \sum_{s=1,2}\bfbeta_s\tp{\bfalpha_s}\zeta(z - \gamma_s) 
       + J\zeta(z) + C, 
\eeqnn
where $C$ is a constant matrix.  By the residue theorem, 
the coefficients of $\zeta(z - \gamma_1)$, $\zeta(z - \gamma_s)$ 
and $\zeta(z)$ have to satisfy the linear equation 
\beqnn
  \sum_{s=1,2}\bfbeta_s\tp{\bfalpha_s} + J = 0 
\eeqnn
that ensures that $A(z)$ is single valued on $\Gamma$.  
Solving this equation for $\bfbeta_s$ leads to 
the formula stated in the lemma.  On the other hand, 
matching with the Laurent expansion of $A(z)$ 
at $z = 0$ leads to the relation 
\beqnn
  A^{(1)} = 
  \sum_{s=1,2}\bfbeta_s\tp{\bfalpha_s}\zeta(- \gamma_s) 
  + C, 
\eeqnn
which determines $C$ as shown in the formula above. 
\qed

The final task is to fulfill the requirement on 
the auxiliary linear system.  By Krichever's lemma 
\cite[Lemma 5.2]{bib:Kr02}, the auxiliary linear system 
$\rd_x\psi(z) = A(z)\psi(z)$ has a $2 \times 2$ 
matrix solution that is holomorphic at $z = \gamma_s$ 
and invertible except at these points if and only if 
$\gamma_s$ and $\alpha_s$ satisfy the equations 
\beq
  \rd_x \gamma_s + \Tr \bfbeta_s\tp{\bfalpha_s} = 0, 
  \label{eq:rd-x-gamma} \\ 
  \rd_x \tp{\bfalpha_s} + \tp{\bfalpha_s}A^{(s,1)} 
  = \kappa_s\tp{\bfalpha_s}, 
  \label{eq:rd-x-alpha}
\eeq
where $A^{(s,1)}$ denotes the constant term of 
the Laurent expansion of $A(z)$ at $z = \gamma_s$, 
\beqnn
  A^{(s,1)} 
  = \lim_{z\to\gamma_s}\left( 
    A(z) - \frac{\bfbeta_s\tp{\bfalpha_s}}{z - \gamma_s}
    \right), 
\eeqnn
and $\kappa_s$ is a constant to be determined by 
the equation itself.

\subsection{Generating functions}

The second stage is to introduce two generating functions 
\beqnn
  \phi(z) = I + \sum_{n=1}^\infty \phi_nz^n, \quad 
  U(z) = J + \sum_{n=1}^\infty U_nz^n 
\eeqnn
as counterparts of $U(\lambda)$ and $\phi(\lambda)$ 
in the case of the usual nonlinear Schr\"odinger hierarchy.  
Recall that the point $\lambda = \infty$ of $\PP^1$ 
corresponds to the origin $z = 0$ of the torus $\Gamma$.  

The first generating function $\phi(z)$ is a Laurent series 
that satisfies the differential equation 
\beq
  \rd_x\phi(z) = A(z)\phi(z) - \phi(z)Jz^{-1}, 
\eeq
where $A(z)$ is understood to be its Laurent expansion 
\beq
  A(z) = Jz^{-1} + \sum_{n=1}^\infty A^{(n)}z^{n-1} 
\eeq
at $z = 0$.  One can construct a solution of 
this differential equation by essentially the same 
(but slightly more complicated) procedure as 
mentioned in the case of the usual nonlinear 
Schr\"odinger  hierarchy.  

The second generating function $U(z)$ can be 
obtained from $\phi(z)$ as 
\beq
  U(z) = \phi(z)J\phi(z)^{-1}, 
\eeq
which satisfies the equations 
\beq
  [\rd_x - A(z),\; U(z)] = 0, \quad 
  U(z)^2 = I. 
\eeq
The Laurent coefficients are again determined 
by a set of recurrence relations: 
\beq
  2JU_{n+1} 
  = \rd_x U_n - \sum_{m=1}^{n+1}[A^{(m)},U_{n+1-m}] 
    - \sum_{m=1}^n U_m U_{n+1-m}. 
\eeq

\subsection{Construction of hierarchy}

The third stage is to construct a set of generators 
$A_n(z)$, $n = 1,2,\ldots$, of time evolutions.  
Just like $A(z)$, they are $2 \times 2$ matrices 
of elliptic functions and characterized 
by the following properties. 
\begin{enumerate}
\item $A_n(z)$ has poles at $z = 0,\gamma_1,\gamma_2$ 
and is holomorphic at other points. 
\item As $z \to 0$, 
\beqnn
  A_n(z) = U(z)z^{-n} + O(z). 
\eeqnn
\item As $z \to \gamma_s$, $s = 1,2$, 
\beqnn
  A_n(z) 
  = \frac{\bfbeta_{n,s}\tp{\bfalpha_s}}{z - \gamma_s} 
    + O(1), 
\eeqnn
where $\bfbeta_{n,s}$ is a two-dimensional column vector 
that does not depend on $z$.  
\end{enumerate}

\begin{lemma}
If $\alpha_1 \not= \alpha_2$, a matrix $A_n(z)$ 
of meromorphic functions on $\Gamma$ 
with these properties does exists.  It is unique 
and can be written explicitly as 
\beq
  A_n(z) 
  &=& \sum_{s=1,2}\bfbeta_{n,s}\tp{\bfalpha_s} 
     (\zeta(z - \gamma_s) + \zeta(\gamma_s)) \\
  && \mbox{} 
  + \sum_{m=0}^{n-1}\frac{(-1)^{m}}{m!}
      \rd_z^{m}\zeta(z)U_{n-1-m} 
  + U_n.  \nonumber 
\eeq
The vectors $\bfbeta_{n,s}$ are determined by 
the linear equation 
\beq
  \sum_{s=1,2}\bfbeta_{n,s}\tp{\bfalpha_s} + U_{n-1} = 0 
\eeq
that ensures the single-valuedness of $A_n(z)$ 
on $\Gamma$.  
\end{lemma}

In the following, the genericity condition
\beq
  \alpha_1 \not= \alpha_2
\eeq
is always assumed; the matrices $A(z)$ and $A_n(z)$ 
are thereby determined.  The elliptic analogue of 
the nonlinear Schr\"odinger hierarchy is defined by 
the Lax equations 
\beq
  [\rd_{t_n} - A_n(z),\; U(z)] = 0 
  \label{eq:Lax-tn-U}
\eeq
for the generating function $U(z)$ 
and the differential equations 
\beq
  \rd_{t_n}\gamma_s + \Tr \bfbeta_{n,s}\tp{\bfalpha_s} = 0, 
  \label{eq:rd-tn-gamma}\\
  \rd_{t_n}\tp{\bfalpha_n} + \tp{\bfalpha_s}A_n^{(s,1)} 
  = \kappa_{n,s}\bfalpha_s 
  \label{eq:rd-tn-alpha}
\eeq
for the Tyurin parameters.  Here $A_n^{(s,1)}$ denotes 
the constant term of the Laurent expansion of $A_n(z)$  
at $z = \gamma_s$, i.e., 
\beqnn
  A_n^{(s,1)} 
  = \lim_{z\to\gamma_s}\left(
      A_n(z) - \frac{\bfbeta_{n,s}\tp{\bfalpha_s}}{z - \gamma_s}
    \right), 
\eeqnn
and $\kappa_{n,s}$ is a constant determined by 
the differential equation itself.  
(\ref{eq:rd-tn-gamma}) and (\ref{eq:rd-tn-alpha}) are 
the necessary and sufficient conditions for the auxiliary 
linear system $\rd_{t_n}\psi(z) = A_n(z)\psi(z)$ to have 
a $2 \times 2$ matrix solution that is holomorphic at 
$z = \gamma_s$ and invertible except at these points. 

One can confirm that the zero-curvature equations 
\beq
  [\rd_{t_m} - A_m(z),\; \rd_{t_n} - A_n(z)] = 0 
  \label{eq:zc-Am-An}
\eeq
are satisfied by any solution of the three equations 
(\ref{eq:Lax-tn-U}), (\ref{eq:rd-tn-gamma}) and 
(\ref{eq:rd-tn-alpha}).  This implies, in particular, 
the commutativity of flows generated by $A_n(z)$.  
Actually, the following stronger statement holds 
as in the case of the usual nonlinear Schr\"odinger hierarchy. 

\begin{theorem}
The system of Lax equations (\ref{eq:Lax-tn-U}) and 
the system of zero-curvature equations (\ref{eq:zc-Am-An}) 
are equivalent under the equations (\ref{eq:rd-tn-gamma}) 
and (\ref{eq:rd-tn-alpha}) for the Tyurin parameters.  
\end{theorem}

As regards the status of (\ref{eq:rd-tn-gamma}) and 
(\ref{eq:rd-tn-alpha}), one can derive them from 
the zero-curvature equations 
\beq
  [\rd_{t_n} - A_n(z),\; \rd_x - A(z)] = 0 
  \label{eq:zc-An-A}
\eeq
assuming that (\ref{eq:rd-x-gamma}) and (\ref{eq:rd-x-alpha}) 
are satisfied.  In this respect, (\ref{eq:rd-x-gamma}) and 
(\ref{eq:rd-x-alpha}) should be understood as part of 
the definition of $A(z)$.  Krichever's construction of 
a hierarchy is rather based on these zero-curvature 
equations \cite{bib:Kr02}.

\section{Elliptic analogue of nonlinear Schr\"odinger hierarchy 
in Grassmannian perspective} 

A technical clue to the Grassmannian perspective of 
the elliptic analogue of the nonlinear Schr\"odinger hierarchy 
is again a factorization or Riemann-Hilbert problem.  
The situation is, however, far more complicated.  
First of all, the present case is concerned with the torus 
rather than the sphere.  Moreover, whereas the usual 
Riemann-Hilbert problem on the sphere is based on the triviality 
of a holomorphic bundle (cf. Section 3.1), the present case is, 
by construction, related to a {\it nontrivial} holomorphic 
bundle in the Tyurin parameterization.  
As it turns out, what is relevant to the present setting 
is a Riemann-Hilbert problem with {\it degeneration points\/}; 
Tyurin parameters are nothing but the geometric data of 
those points.  This kind of Riemann-Hilbert problems 
also appear in the work of Krichever and Novikov 
\cite{bib:Kr78,bib:KN78,bib:KN80} on commutative rings 
of differential operators.  

Another clue can be found in the paper of Previato 
and Wilson \cite{bib:PW89}.  They demonstrate therein 
a ``dressing method'' based on an infinite dimensional 
Grassmann variety to solve a Riemann-Hilbert problem 
of the Krichever-Novikov type.  Moreover, their paper 
shows what should be the ``vacuum'' that corresponds 
to a holomorphic vector bundle in the Tyurin parametrization.  

These ideas lead to a Grassmannian perspective 
of the elliptic analogue \cite{bib:Ta03a}.

\subsection{Riemann-Hilbert problem with degeneration points} 

In the following, $t$ denotes the full set 
of time variables $(t_1,t_2,\ldots)$, 
in which $x$ is identified with $t_1$.  
Moreover, any quantity that depends on $t$ 
is written with its $t$-dependence indicated explicitly  
as $A(t,z)$, $A_n(t,z)$, $\gamma_s(t)$, $\alpha_s(t)$, etc.  

The Lax equations (\ref{eq:Lax-tn-U}) and 
the zero-curvature equations (\ref{eq:zc-Am-An}) 
are associated with the auxiliary linear system 
\beqnn
  \rd_{t_n}\psi(t,z) = A_n(t,z)\psi(t,z). 
\eeqnn
The Riemann-Hilbert problem is concerned with 
two distinct solutions of this linear system. 

One solution is the Laurent series solution of 
the form 
\beq
  \psi(t,z) 
  = \phi(t,z)\exp\Bigl(\sum_{n=1}^\infty t_nJz^{-n}\Bigr), 
\eeq
where the prefactor $\phi(t,z)$ is a Laurent series of 
the form 
\beqnn
  \phi(t,z) = I + \sum_{n=1}^\infty \phi_n(t)z^n. 
\eeqnn
This prefactor is nothing but the generating function 
introduced previously, but it is now required to 
satisfy the differential equations
\beq
  \rd_{t_n}\phi(t,z) = A_n(t,z)\phi(t,z) - \phi(t,z)Jz^{-n} 
\eeq
for $n = 1,2,\ldots$ as well.  Note that 
this Laurent series solution, by its nature, 
carries no information on the global structure 
of $A_n(t,z)$'s.  

Another solution $\chi(t,z)$ is characterized by 
the initial condition 
\beq
  \chi(0,z) = I. 
\eeq
This solution $\chi(t,z)$ turns out to carry 
global information.  To avoid delicate problems, 
suppose that the solutions of the hierarchy under 
consideration are (real or complex) analytic 
in a neighborhood of the initial point $t = 0$.  
One can then expand it to a Taylor series in $t$. 
The Taylor coefficients of $\chi(t,z)$ at $t = 0$ 
can be evaluated by successively differentiating 
the differential equations as 
\beqnn
  \rd_{t_n}\chi(t,z) 
    &=& A_n(t,z)\chi(t,z), 
  \nonumber \\
  \rd_{t_m}\rd_{t_n}\chi(t,z) 
    &=& (\rd_{t_m}A_n(t,z) + A_n(t,z)A_m(t,z))\chi(t,z), 
  \nonumber \\
  \rd_{t_k}\rd_{t_m}\rd_{t_n} \chi(t,z) 
    &=& \Bigl(\rd_{t_k}\rd_{t_m}A_n(t,z) 
          + \rd_{t_k}(A_n(t,z)A_m(t,z)) 
    \nonumber \\
    && \mbox{} 
          + (\rd_{t_m}A_n(t,z))A_k(t,z) 
          + A_n(t,z)A_m(t,z)A_k(t,z)\Bigr)\chi(t,z), 
\eeqnn
etc.   Letting $t = 0$, we are left with 
a polynomial of derivatives of $A_n$'s.  
One can deduce from these calculations that 
all Taylor coefficients of $\chi(t,z)$ at $t = 0$ 
are a matrix of meromorphic functions of $z$ on 
$\Gamma$ with poles at $z = 0,\gamma_1(0),\gamma_2(0)$ 
and holomorphic at other points.  Since the order 
of poles at $z = 0$ is unbounded for higher orders 
of the Taylor expansion, the Taylor series of 
$\chi(t,z)$ has an essential singularity 
at $z = 0$.  On the other hand, the poles 
at $z = \gamma_1(0),\gamma_2(0)$ remain to be 
of the first order.  More careful analysis 
\cite{bib:Ta03a} shows that the detailed structure 
of these first order poles: 

\begin{lemma}
As $z \to \gamma_s(0)$, $s = 1,2$, $\chi(t,z)$ behaves as 
\beqnn
  \chi(t,z) 
  = \frac{\bfbeta_{\chi,s}(t)\tp{\bfalpha_s(0)}}
         {z - \gamma_s(0)} 
    + O(1), 
\eeqnn
where $\bfbeta_{\chi,s}(t)$ is a two-dimensional vector. 
\end{lemma}

Another important property of $\chi(t,z)$ can be 
seen from the the linear system 
\beqnn
  \rd_x\chi(t,z) = A(t,z)\chi(t,z). 
\eeqnn
Taking the residue at $z = \gamma_s(t)$ yields 
the relation 
\beqnn
  0 = \bfbeta_s(t)\tp{\bfalpha_s(t)}\chi(t,\gamma_s(t)), 
\eeqnn
which implies that 
\beqnn
  \tp{\bfalpha_s(t)}\chi(t,\gamma_s(t)) = \mathbf{0}. 
\eeqnn
Thus one finds the following. 

\begin{lemma} $\det\chi(t,z)$ has zeroes at 
$z = \gamma_s(t)$, $s = 1,2$.  $\tp{\bfalpha_s(t)}$ 
is a left null vector of $\chi(t,\gamma_s(t))$.  
\end{lemma}

This result shows that $\chi(t,z)$ is exactly the solution 
mentioned in Krichever's lemma \cite[Lemma 5.2]{bib:Kr02}, 
namely a matrix solution of the auxiliary linear system 
that is holomorphic at the movable poles of $A(t,z)$.   

Since $\psi(t,z)$ and $\chi(t,z)$ satisfy the same 
auxiliary linear system, their ``matrix ratio'' 
$\chi(t,z)^{-1}\psi(t,z)$ is independent of $t$, 
hence equal to its initial value at $t = 0$.  
One thus obtains the relation 
\beq
  \psi(0,z) 
  = \chi(t,z)^{-1}\psi(t,z) 
\eeq
or, equivalently, 
\beq
  \phi(0,z) \exp\Bigl(- \sum_{n=1}^\infty t_nJz^{-n}\Bigr) 
  = \chi(t,z)^{-1}\phi(t,z).  
\eeq
This is the Riemann-Hilbert problem that plays 
the role of an intermediate step towards 
the Grassmannian perspective.  The pair of 
$\phi(t,z)$ and $\chi(t,z)$ are referred to 
as a Riemann-Hilbert pair.  

Note that this Riemann-Hilbert problem has a few 
unusual aspects.  Firstly, in addition to the pole 
at $z = 0$, $\chi(t,z)$ has extra poles at 
$z = \gamma_s(0)$, $s = 1,2$.  Secondly, $\chi(t,z)$ 
{\it degenerate} (namely, $\det\chi(t,z)$ has zeroes) 
at $z = \gamma_s(t)$, $s = 1,2$.  Moreover, 
these degeneration points are movable as $t$ varies.

\subsection{Grassmann variety and vacuum} 

Let $V$ denote the vector space of all $2 \times 2$ 
matrices of Laurent series 
\beq
  X(z) = \sum_{n=-\infty}^\infty X_nz^n, \quad 
  X_n \in \mathrm{gl}(2,\CC), 
\eeq
that converges in a neighborhood of $z = 0$ 
except at $z = 0$, and $V_{+}$ the subspace 
\beq
  V_{+} = \{X(z) \in V \mid  \mbox{$X_n = 0$ for $n \le 0$}\}. 
\eeq
of all $X(z) \in V$ that are holomorphic and vanish at 
$z = 0$.  Recalling that $z$ amounts to $\lambda^{-1}$, 
this is essentially the same setting as the case of 
the nonlinear Schr\"odinger hierarchy.  
The Grassmann variety $\mathrm{Gr}$ and the big cell 
$\mathrm{Gr}^\circ \subset \mathrm{Gr}$ are defined as 
\beq
\lefteqn{\mathrm{Gr} = \{W \subset V \mid} \\
  && \dim\Ker(W \to V/V_{+}) 
     = \dim\Coker(W \to V/V_{+}) < \infty\} 
  \nonumber 
\eeq
and 
\beq
  \mathrm{Gr}^\circ = \{W \in \mathrm{Gr} \mid W \simeq V/V_{+}\}. 
\eeq

The following lemma shows the construction of a special 
point $W_0(\gamma,\alpha)$ of $\mathrm{Gr}^\circ$, 
which plays the role of ``vacuum'' in the present setting.  
This is a matrix version of the vacuum that Previato 
and Wilson \cite{bib:PW89} suggest to use for a holomorphic 
vector bundle in the Tyurin parametrization.  

\begin{lemma}\label{lem:W0-basis}
Let $\gamma = (\gamma_1,\gamma_2)$ be a pair of distinct 
points of $\Gamma$, $\gamma_1 \not= \gamma_2$, and 
$\alpha = (\alpha_1,\alpha_2)$ a pair of constants 
satisfying the genericity condition 
$\alpha_1 \not= \alpha_2$.  Then, 
for any integer $n \ge 0$ and the matrix indices 
$i,j = 1,2$, there is a unique $2 \times 2$ matrix 
$w_{n,ij}(z)$ of meromorphic functions on $\Gamma$ 
with the following properties: 
\begin{enumerate}
\item $w_{n,ij}(z)$ has poles at 
$z = 0,\gamma_1,\gamma_2$ and is holomorphic 
at other points.  
\item $w_{n,ij}(z) = E_{ij}z^{-n} + O(z)$ as $z \to 0$, 
where $E_{ij}$, $i,j = 1,2$, are the standard basis 
of $\mathrm{gl}(2,\CC)$.  
\item As $z \to \gamma_s$, $s = 1,2$, 
\beqnn
  w_{n,ij}(z) 
  = \frac{\bfbeta_{n,ij,s}\tp{\bfalpha_s}}
         {z - \gamma_s} 
    + O(1), 
\eeqnn
where $\bfalpha_s = \tp{(\alpha_s,1)}$, and 
$\bfbeta_{n,ij,s}$ is another two-dimensional 
constant vector. 
\end{enumerate}
The subspace 
\beq
  W_0(\gamma,\alpha) 
  = \langle w_{n,ij}(z) \mid n \ge0,\; i,j = 1,2 \rangle 
\eeq
spanned by (the Laurent series of) $w_{n,ij}(z)$'s 
is an element of the big cell. 
\end{lemma}

This vacuum $W_0(\gamma,\alpha)$ is ``dressed'' by 
a Laurent series to become a dressed vacuum: 
\beqnn
  W = W_0(\gamma,\alpha) \phi(z), \quad 
  \phi(z) = I + \sum_{n=1}^\infty \phi_n z^n, \quad 
  \phi_n \in \mathrm{gl}(2,\CC). 
\eeqnn
The set 
\beq
\lefteqn{\mathcal{M} = \{ W \in \mathrm{Gr}^o \mid 
      W = W_0(\gamma,\alpha)\phi(z),\;
      \phi_n \in \mathrm{gl}(2,\CC),} \\ 
  &&  \gamma = (\gamma_1,\gamma_2) \in \Gamma^2, \; 
      \alpha = (\alpha_1,\alpha_2) \in \CC^2,\; 
      \gamma_1 \not= \gamma_2, \; 
      \alpha_1 \not= \alpha_2       
      \} 
  \nonumber
\eeq
of these dressed vacua is the phase space for 
the Grassmannian perspective of the elliptic analogue 
of the nonlinear Schr\"odinger hierarchy.

\subsection{Interpretation of Riemann-Hilbert problem}

For technical reasons, the following consideration 
is limited to a small neighborhood of $t = 0$.  
The goal is to translate the Riemann-Hilbert problem 
to the language of the set $\mathcal{M}$ of dressed vacua. 
A clue is the the following. 

\begin{lemma}
$W_0(\gamma(t),\alpha(t))\chi(t,z) = W_0(\alpha(0),\gamma(0))$. 
\end{lemma}

\proof 
The following is an outline of the proof; see the paper 
\cite{bib:Ta03a} for details.  Let $w_{n,ij}(t,z)$, 
$n \ge 0$, $i,j=1,2$, denote the elements of the basis 
of $W_0(\gamma(t),\alpha(t))$ defined in Lemma 
\ref{lem:W0-basis}.  $w_{n,ij}(t,z)$ has poles at 
$z = 0,\gamma_1(t),\gamma_2(t)$, and behaves as 
\beqnn
  w_{n,ij}(t,z) 
  = \frac{\bfbeta_{n,ij,s}(t)\tp{\bfalpha_s(t)}}{z - \gamma_s(t)} 
    + O(1) 
\eeqnn
as $z \to \gamma_s(t)$.  Upon multiplication 
with $\chi(t,z)$, the poles at $z = \gamma_s(t)$ 
are cancelled out because $\tp{\bfalpha_s(t)}$ is 
a left null vector of $\chi(t,\gamma_s(t))$.  
Thus $w_{n,ij}(t,z)\chi(t,z)$ turns out to 
have an essential singularity at $z = 0$, 
first order poles at $z = \gamma_s(0)$, $s = 1,2$,  
and is holomorphic at other points.  The leading part 
of the Laurent expansion at $z = \gamma_s(0)$ takes 
the form 
\beqnn
  w_{n,ij}(t,z)\chi(t,z) 
  = \frac{w_{n,ij}(t,\gamma_s(0))\bfbeta_{\chi,s}(t)
      \tp{\bfalpha_s(0)}}{z - \gamma_s(0)} 
    + O(1). 
\eeqnn
These results show that $w_{n,ij}(t,z)\chi(t,z)$ 
is an element of $W_0(\gamma(0),\alpha(0))$.  
One can thus see that 
\beqnn
  W_0(\gamma(t),\alpha(t))\chi(t,z) \subseteq 
  W_0(\gamma(0),\alpha(0)).
\eeqnn
A few more steps of consideration on the analytic 
properties of $\chi(t,z)$ lead to the conclusion 
that these two vector subspaces of $V$ are equal.  
\qed 

Thanks to this lemma, one can readily convert 
the Riemann-Hilbert problem to the language of 
dressed vacua.  The Riemann-Hilbert relation 
yields the relation 
\beqnn
  W_0(\gamma(t),\alpha(t))\phi(t,z) 
  = W_0(\gamma(t),\alpha(t))\chi(t,z)\phi(0,z) 
    \exp\Bigl(- \sum_{n=1}^\infty t_nJz^{-n}\Bigr). 
\eeqnn
The lemma shows that $W_0(\gamma(t),\alpha(t))$ 
absorbs $\chi(t,z)$ to become $W_0(\gamma(0),\alpha(0))$. 
The outcome is the relation 
\beqnn
  W_0(\gamma(t),\alpha(t))\phi(t,z) 
  = W_0(\gamma(0),\alpha(0))\phi(0,z) 
    \exp\Bigl(- \sum_{n=1}^\infty t_nJz^{-n}\Bigr), 
\eeqnn
which means that the dressed vacuum 
$W(t) = W_0(\gamma(t),\alpha(t))\phi(t,z) \in \mathcal{M}$ 
obeys the exponential law 
\beq
  W(t) = W(0)\exp\Bigl(- \sum_{n=1}^\infty t_nJz^{-n}\Bigr). 
\eeq

Conversely, one can derive a solution of 
the Riemann-Hilbert problem from these exponential flows 
as follows.  (This is a variation of the dressing method 
of Previato and Wilson \cite{bib:PW89}.)  Given a set of 
initial values $\gamma(0), \alpha(0)$ and $\phi(0,z)$, 
one can consider the exponential flows sending $W(0) 
= W_0(\gamma(0),\alpha(0))\phi(0,z)$ to $W(t)$. 
If $t$ is sufficiently small, $W(t)$ remains 
in the big cell. This means that the linear map 
$W(t) \to V/V_{+}$ is an isomorphism.  
Let $\phi(t,z) \in W(t)$ be the inverse image of 
$I \in V/V_{+}$ by this isomorphism.  
Being equal to $I$ modulo $V_{+}$, $\phi(t,z)$ is 
a Laurent series of the form 
\beqnn
  \phi(t,z) = 1 + \sum_{n=1}^\infty \phi_n(t)z^n. 
\eeqnn
On the other hand, as an element of 
\beqnn
  W(t) = W_0(\gamma(0),\alpha(0))\phi(0,z)
    \exp\Bigl(- \sum_{n=1}^\infty t_nJz^{-n}\Bigr), 
\eeqnn
$\phi(t,z)$ can also be expressed as 
\beqnn
  \phi(t,z) = \chi(t,z)\phi(0,z)
    \exp\Bigl(- \sum_{n=1}^\infty t_nJz^{-n}\Bigr) 
\eeqnn
with an element $\chi(t,z)$ of $W_0(\gamma(0),\alpha(0))$.  
Thus one obtains a Riemann-Hilbert pair. The associated 
Tyurin parameters $(\gamma_s(t),\bfalpha_s(t))$ are 
determined as the position of zeros of $\chi(t,z)$ 
and the normalized left null vector of $\chi(t,z)$ at 
those degeneration points.  

One thus eventually arrives at the following 
Grassmannian perspective in the present setting. 

\begin{theorem}
The elliptic analogue of the nonlinear Schr\"odinger hierarchy 
can be mapped, by the correspondence 
$W(t) = W_0(\gamma(t),\alpha(t))\phi(t,z)$, to a dynamical 
system on the set $\mathcal{M}$ of dressed vacua in 
the Grassmann variety $\mathrm{Gr}$.  The motion of $W(t)$ 
obeys the exponential law.  Conversely, the exponential flows 
on $\mathcal{M}$ yield a solution of the Riemann-Hilbert problem. 
\end{theorem}

As a final remark, it should be stressed that the main 
characters of this story are all related to the geometry 
of holomorphic vector bundles over $\Gamma$.  
The Tyurin parameters $(\gamma(t),\alpha(t))$ correspond 
to a holomorphic vector bundle that deforms as $t$ varies.  
The subspace $W_0(\gamma,\alpha) \subset V$ can be 
identified with the space of holomorphic sections of 
the associated $\mathrm{sl}(2,\CC)$ bundle over 
the punctured torus $\Gamma \setminus \{z = 0\}$.  
$\phi(t,z)$ is related to changing local trivialization 
of this bundle at $z = 0$.  Note, in particular, 
that the primary role (as a dynamical variable) is now played 
by the data of local trivialization.  This differs decisively 
from the work of Previato and Wilson \cite{bib:PW89}; they take, 
in place of the data of local trivialization, a set of functions 
in Krichever's ``algebraic spectral data'' \cite{bib:Kr78} 
as main parameters.  In this respect, the present setting is 
rather close to Li and Mulase's approach \cite{bib:Mu90,bib:LM97} 
to commutative rings of differential operators; they treat 
the choice of local trivialization as an independent data.

\section{Landau-Lifshitz hierarchy in Grassmannian perspective} 

The last example with an elliptic spectral parameter is 
the Landau-Lifshitz equation in $1 + 1$ dimensions and 
the associated hierarchy (Landau-Lifshitz hierarchy) of 
higher time evolutions.  This is one of the classical 
examples of soliton equations with an elliptic 
zero-curvature  representation \cite{bib:Sk79,bib:Ch81}.  

As regards the Grassmannian perspective of this equation, 
studies from a very close point of view have been done 
by Date, Jimbo, Kashiwara and Miwa \cite{bib:DJKM83} 
and Carey, Hannabuss, Mason and Singer \cite{bib:CHMS93}. 
Actually, Date et al. developed a free fermion formalism 
rather than a Grassmannian formalism.  Carey et al. 
presented two approaches to a factorization method 
for solving the Landau-Lifshitz equation.  The first 
approach uses an infinite dimensional Grassmann 
manifold (rather than a ``variety'', because this is 
a functional analytic model).  The second one is based 
on the geometry of a holomorphic vector bundle over 
the torus $\Gamma = \CC/(2\omega_1\ZZ + 2\omega_3\ZZ)$.  
This work is yet unsatisfactory because their usage 
of the Grassmann manifold fails to incorporate 
the bundle structure.  

The lessons in the preceding examples show that 
a clue is always the choice of a suitable ``vacuum'' 
(and of course a Grassmann variety that accommodates 
that vacuum).  As the paper of Previato and Wilson 
suggests \cite{bib:PW89}, a correct choice of vacuum 
is somehow related to the structure of a holomorphic 
vector bundle.  The Grassmannian perspective of 
the Landau-Lifshitz equation (and hierarchy), too, 
can be reached along the same lines \cite{bib:Ta03b}.

\subsection{Geometric and algebraic structures 
behind Landau-Lifshitz equation}

The zero-curvature representation of the Landau-Lifshitz 
equation \cite{bib:Sk79,bib:Ch81} is based on 
the first order matrix differential operator 
$\rd_x - A(z)$ with the $A$-matrix of the form 
\beq
  A(z) = \sum_{a=1,2,3} w_a(z)S_a\sigma_a, 
\eeq
where $S_a$'s are dynamical variables (spin fields) 
and $\sigma_a$'s denote the Pauli matrices. 
The weight functions $w_a(z)$ are defined by Jacobi's 
elliptic functions $\mathrm{sn},\mathrm{cn},\mathrm{dn}$ as
\beq
  w_1(z) = \frac{\alpha\mathrm{cn}(\alpha z)}{\mathrm{sn}(\alpha z)}, 
  \quad 
  w_2(z) = \frac{\alpha\mathrm{dn}(\alpha z)}{\mathrm{sn}(\alpha z)}, 
  \quad 
  w_3(z) = \frac{\alpha}{\mathrm{sn}(\alpha z)}, 
\eeq
where $\alpha = \sqrt{e_1 - e_3}$, $e_a = \wp(\omega_a)$. 

The matrix $A(z)$ has the twisted double periodicity 
\beq
  A(z + 2\omega_a) = \sigma_a A(z) \sigma_a, 
  \quad a = 1,2,3, 
\eeq
where $\omega_2$ denotes the third half period 
$\omega_2 = - \omega_1 - \omega_3$.  This is 
a manifestation of the structure of a nontrivial 
holomorphic $\mathrm{sl}(2,\CC)$ bundle over the torus;  
$A(z)$ is a meromorphic section of that bundle.  
The same bundle is known to play a fundamental role 
in the elliptic Gaudin model and an associated 
conformal field theory \cite{bib:KT97}. 

Compared with the equations formulated by Tyurin parameters, 
the Landau-Lifshitz equation is rather close to classical 
soliton equations with a rational zero-curvature 
representation, because one can treat this system 
by a factorization method based on a Lie group of 
Laurent series (or a loop group) with factorization 
structure \cite{bib:RS86,bib:CHMS93}.  Geometrically, 
this fact is related to {\it rigidity} of 
the aforementioned holomorphic $\mathrm{sl}(2,\CC)$ bundle 
or of an associated $\mathrm{SL}(2,\CC)$ bundle 
\cite{bib:HM01}.  

To formulate the factorization structure, 
one starts from a Lie algebra with direct sum 
decomposition to two subalgebras.  
Let $\fkg$ be the Lie algebra of Laurent series 
\beq
  X(z) = \sum_{n=-\infty}^\infty X_nz^n, \quad 
  X_n \in \mathrm{sl}(2,\CC), 
\eeq
that converge in a neighborhood of $z = 0$ except at $z = 0$. 
This Lie algebra has a direct sum decomposition of the form 
\beq
  \fkg = \fkg_{\mathrm{out}} \oplus \fkg_{\mathrm{in}}, 
\eeq
where $\fkg_{\mathrm{in}}$ and $\fkg_{\mathrm{out}}$ 
are the following subalgebras: 
\begin{enumerate}
\item $\fkg_{\mathrm{in}}$ consists of all $X(z) \in \fkg$ 
that are also holomorphic at $z = 0$, i.e, $X_n = 0$ 
for $n < 0$. 
\item $\fkg_{\mathrm{out}}$ consists of all $X(z) 
\in \fkg$ that can be extended to a holomorphic mapping 
$X: \CC \setminus (2\omega_1\ZZ + 2\omega_3\ZZ 
\to \mathrm{sl}(2,\CC)$ with singularity at each point 
of $2\omega_1\ZZ + 2\omega_3\ZZ$ and satisfy 
the twisted double periodicity condition 
\beqnn
  X(z + 2\omega_a) = \sigma_a X(z) \sigma_a 
  \quad a = 1,2,3. 
\eeqnn
\end{enumerate}
Note that constant matrices are excluded from 
$\fkg_{\mathrm{out}}$, so that 
$\fkg_{\mathrm{out}} \cap \fkg_{\mathrm{in}} = \{0\}$. 
One can choose $\{\rd_z^nw_a(z)\sigma_a \mid 
n \ge 0, \; a = 1,2,3\}$ as a basis of $\fkg_{\mathrm{out}}$; 
the projection $(\cdot)_{\mathrm{out}}: 
\fkg \to \fkg_{\mathrm{out}}$  thereby takes the simple form 
\beq
  \left(z^{-n-1}\sigma_a\right)_{\mathrm{out}} 
  = \frac{(-1)^n}{n!}\rd_z^nw_a(z) \sigma_a, \quad 
  \left(z^n\sigma_a\right)_{\mathrm{out}} = 0, \quad
  n \ge 0. 
\eeq

The direct sum decomposition of the Lie algebra $\fkg$ 
induces the factorization of the associated Lie group 
$G = \exp\fkg$ to the subgroups 
$G_{\mathrm{out}} = \exp\fkg_{\mathrm{out}}$ and 
$G_{\mathrm{in}} = \exp\fkg_{\mathrm{in}}$, namely, 
any element $g(z)$ of $G$ near the unit matrix $I$  
can be uniquely factorized as 
\beq
  g(z) = g_{\mathrm{out}}(z)^{-1}g_{\mathrm{in}}(z), \quad 
  g_{\mathrm{out}}(z) \in G_{\mathrm{out}}, \quad 
  g_{\mathrm{in}}(z) \in G_{\mathrm{in}}. 
\eeq

\subsection{Construction of hierarchy}

The Landau-Lifshitz hierarchy can be obtained by 
the projection of the exponential flows 
\beq
  g(\lambda) \mapsto 
  g(\lambda)\exp\Bigl(- \sum_{n=1}^\infty t_nJ\lambda^n\Bigr) 
\eeq
on $G$ to $G_{\mathrm{in}}$ with regard to 
the foregoing factorization \cite{bib:GM91,bib:CHMS93}.  
The fundamental dynamical variable is thus 
a Laurent series of the form 
\beqnn
  \phi(z) = \sum_{n=0}^\infty \phi_nz^n, \quad 
  \det \phi(z) = 1, 
\eeqnn
that converges in a neighborhood of $z = 0$.  
The time evolution $\phi(0,z) \mapsto \phi(t,z)$ 
is achieved by the factorization 
\beq
  \phi(0,z)\exp\Bigl(- \sum_{n=1}^\infty t_nz^{-n}\sigma_3\Bigl) 
  = \chi(t,z)^{-1}\phi(t,z), 
  \label{factor-phi-exp}
\eeq
where $\chi(t,z)$ is an element of $G_{\mathrm{out}}$ 
that also depends on $t$.  As demonstrated in the case 
of the usual nonlinear Schr\"odinger hierarchy, 
one can derive the equations 
\beq
  \rd_{t_n}\phi(t,z) 
  = A_n(t,z)\phi(t,z) - \phi(t,z)z^{-n}\sigma_3, 
\eeq
where 
\beq
  A_n(t,z) = \Bigl(\phi(t,z)z^{-n}\sigma_3\phi(t,z)^{-1}
             \Bigr)_{\mathrm{out}}, 
\eeq
or 
\beq
  \rd_{t_n}\phi(t,z) 
  = - \Bigl(\phi(t,z)z^{-n}\sigma_3\phi(t,z)^{-1}
      \Bigr)_{\mathrm{in}}\phi(t,z) 
\eeq
as equations of motion of $\phi(t,z) \in G_{\mathrm{in}}$.  
$(\cdot)_{\mathrm{in}}$ denotes the projection 
$\fkg \to \fkg_{\mathrm{in}}$. 

The zero-curvature equations 
\beq
  [\rd_{t_m} - A_m(t,z),\; \rd_{t_n} - A_n(t,z)] = 0 
\eeq
follow from the auxiliary linear system 
\beq
  (\rd_{t_n} - A_n(t,z))\chi(t,z) = 0 
\eeq
as the Frobenius integrability condition.

\subsection{Grassmann variety and vacuum} 

It will be reasonable to use the same pair $(V,V_{+})$ 
of vector spaces as those for the elliptic analogues 
of the nonlinear Schr\"odinger hierarchy. Actually, 
already at this stage, the present approach differs 
from that of Carey et al. \cite{bib:CHMS93}.  
Carey et al. use a vector space of two-component vectors 
rather than $2 \times 2$ matrices;  this is not suited 
for treating the aforementioned $\mathrm{sl}(2,\CC)$ 
bundle structure.  

The next problem is the choice of a suitable subspace 
$W_0 \subset V$ that plays the role of ``vacuum.''  
In view of the previous examples, $W_0$ should be 
a vector subspace that absorbs the first factor 
$\chi(t,z)$ of the factorization pair.  This will be 
the case if $W_0$ consists of matrix-valued functions 
of $z$ with the same analytic properties as $\chi(t,z)$.  
As a $t$-dependent element of $G_{\mathrm{out}}$, 
$\chi(t,z)$ is a matrix-valued holomorphic function 
on $\CC \setminus (2\omega_1\ZZ + 2\omega_3\ZZ)$ 
with twisted double periodicity.  

For this reason, let $W_0$ be the subspace of $V$ that
consists of all $X(z) \in V$ with the following properties: 
\begin{enumerate}
\item $X(z)$ can be extended to a holomorphic mapping
\beqnn
  X: \CC \setminus (2\omega_1\ZZ + 2\omega_3\ZZ) 
  \to \mathrm{gl}(2,\CC). 
\eeqnn
\item $X(z)$ has the twisted double periodicity 
\beqnn
  X(z + 2\omega_a) = \sigma_a X(z) \sigma_a, 
  \quad a = 1,2,3. 
\eeqnn
\end{enumerate}
This resembles the definition of $\fkg_{\mathrm{out}}$; 
the difference is, firstly, that $X(z)$ now takes values 
in $\mathrm{gl}(2,\CC)$ rather than $\mathrm{sl}(2,\CC)$, 
and secondly, that $X(z)$ can be a constant matrix.  

As it turns out, this subspace $W_0$ does {\it not} 
satisfy the condition in the definition of 
the Grassmann variety $\mathrm{Gr}$ that has been used 
in the previous case: 

\begin{lemma}
The following hold for the linear map $W_0 \to V/V_{+}$: 
\begin{enumerate}
\item 
$\Im(W_0 \to V/V{+})\oplus \mathrm{sl}(2,\CC) 
\oplus \CC z^{-1}I = V/V_{+}$. 
\item
$\Ker(W_0 \to V/V_{+}) = \{0\}$. 
\end{enumerate}
\end{lemma}

\proof 
It is a (slightly advanced) exercise of linear algebra 
and complex function theory to confirm that $W_0$ is 
spanned by $I$, $\rd_z^nw_a(z)\sigma_a$, $a = 1,2,3$, 
and $\rd_z^n\wp(z)I$ for $n \ge 0$.  
$\rd_z^nw_a(z)$ and $\rd_z^n\wp(z)$ have 
the Laurent expansion 
\beqnn
  \rd_z^nw_a(z)\sigma_a = (-1)^n n!z^{-n-1}\sigma_a + O(z) 
\eeqnn
and 
\beqnn
  \rd_z^n\wp(z)I = (-1)^n (n+1)!z^{-n-2}I + O(z) 
\eeqnn
at $z = 0$.  This implies that these generators 
of $W_0$ are linearly independent, and that 
the image of $W \to V/V_{+}$ are spanned by 
$I$, $z^{-n-1}\sigma_a$, $a = 1,2,3$, and 
$z^{-n-2}I$ for $n \ge 0$ among the standard basis 
$\{z^{-n}\sigma_a,\, z^{-n}I \mid n \ge 0, \; a = 1,2,3\}$ 
of $V/V_{+}$.  What is missing are $\sigma_a$, $a = 1,2,3$, 
and $z^{-1}I$, which respectively span the subspaces 
$\mathrm{sl}(2,\CC)$ and $\CC z^{-1}$ of $V/V_{+}$.   
Thus the assertion on $\Im(W_0 \to V/V_{+})$ follows.  
On the other hand, one has $\Ker(W_0 \to V/V_{+}) 
= W_0 \cap V_{+}$.  Any element $X(z)$ of $W_0 \cap V_{+}$ 
has the twisted double periodicity and a zero at 
all points of $2\omega_1\ZZ + 2\omega_3\ZZ$; 
by Liouville's theorem, such a matrix-valued function 
is identically zero. 
\qed

This lemma implies that 
\beq
  \dim\Ker(W_0 \to V/V_{+}) = 0, \quad 
  \dim\Coker(W_0 \to V/V_{+}) = 4. 
\eeq
Consequently, the Grassmann variety to accommodate 
$W_0$ is not $\mathrm{Gr}$ but the following one: 
\beq
\lefteqn{\mathrm{Gr}_{-4} = \{ W \subset V \mid} \\
  && \dim\Ker(W \to V/V_{+}) 
     = \dim\Coker(W \to V/V_{+}) - 4 < \infty \}. 
  \nonumber 
\eeq
The subset 
\beq
  \mathrm{Gr}_{-4}^\circ 
  = \{ W \in \mathrm{Gr}_{-4} \mid 
  W \simeq V/(V_{+}\oplus\mathrm{sl}(2,\CC)\oplus\CC z^{-1}I) \} 
\eeq
of $\mathrm{Gr}_{-4}$ is an open subset, in fact, 
the open cell (or ``big cell'') of a cell decomposition 
of $\mathrm{Gr}_{-4}$.  The foregoing lemma shows 
that $W_0$ is actually an element of this open subset: 
\beq
  W_0 \in \mathrm{Gr}_{-4}^\circ. 
\eeq
The set 
\beq 
  \mathcal{M} 
  = \{ W \in \mathrm{Gr}_{-4}^\circ \mid 
    W = W_0\phi(z), \; \phi(z) \in G_{\mathrm{in}} \} 
\eeq
of dressed vacua becomes the phase space of 
a dynamical system to which the Landau-Lifshitz 
hierarchy is mapped.

\subsection{Interpretation of factorization problem}

The following consideration is, again, 
limited to a small neighborhood of $t = 0$.  
In this situation, one can prove the following 
in the same way as the case of the nonlinear 
Schr\"odinger hierarchy. 

\begin{lemma}
$W_0\chi(t,z) = W_0$.  
\end{lemma}

Using this lemma, one can repeat 
the calculations done for the previous cases 
to show that the motion of the dressed vacuum 
$W(t) = W_0\phi(t,z) \in \mathcal{M}$ obeys 
the exponential law 
\beq
  W(t) = W(0)\exp\Bigl(- \sum_{n=1}^\infty t_nJz^{-n}\Bigr). 
\eeq

The converse, namely, deriving a solution of 
the factorization problem from the exponential flows 
needs an extra effort because the definition of 
the big cell is different from the previous cases.  
This is also related to the fact that the leading 
term $\phi_0(t)$ of $\phi(t,z)$ is generally not equal 
to $I$.  A clue here is the fact that $W(t)$, as 
an element of the big cell, satisfies the condition that 
\beq
  \dim\Im(W(t)\to V/V_{+}) \cap \mathrm{gl}(2,\CC) = 1. 
\eeq
The leading term $\phi_0(t)$ is picked out from 
this one dimensional subspace; if $t$ is sufficiently small, 
$\phi_0(t)$ is an invertible matrix.  The rest of 
the construction is almost parallel to the case 
of the nonlinear Schr\"odinger hierarchy; 
see the paper \cite{bib:Ta03b} for details.  

The conclusion is that the Grassmannian perspective 
also holds for this case, but with a different 
Grassmann variety:  

\begin{theorem}
The Landau-Lifshitz hierarchy can be mapped, 
by the correspondence $W(t) = W_0 \phi(t,\lambda)$, 
to a dynamical system on the set $\mathcal{M}$ of 
dressed vacua in the Grassmann variety $\mathrm{Gr}_{-4}$.  
The motion of $W(t)$ obeys the exponential law. 
Conversely, the exponential flows on $\mathcal{M}$ 
yield a solution of the factorization problem. 
\end{theorem}

\section{Conclusion} 

A main conclusion of this case study is that 
the structure of a holomorphic vector bundle is 
the most important clue to the Grassmannian 
perspective of soliton equations with a zero-curvature 
representation constructed on an algebraic curve. 
This is also the case for classical soliton equations 
with a rational spectral parameter;  the relevant 
holomorphic vector bundle therein is a trivial bundle.  
The two examples with an elliptic spectral parameters 
examined here are respectively accompanied by a bundle 
of its own particular type.  The bundle for the elliptic 
nonlinear Schr\"odinger hierarchy is naturally the one 
in the Tyurin parametrization.  The bundle for 
the Landau-Lifshitz hierarchy is a rigid bundle.  

It is remarkable that the mapping to an infinite dimensional 
Grassmann variety can be constructed in a fully parallel, 
almost universal way.  Namely, the first thing to do 
is to choose a special base point $W_0$, called  ``vacuum,'' 
of the Grassmann variety.  This is determined by the relevant 
holomorphic vector bundle $E$.  More precisely, one has to 
choose a marked point $P_0$ of $\Gamma$, a local coordinate $z$ 
in a neighborhood of $P_0$ and a local trivialization of $E$ 
in a neighborhood of $P_0$ as extra geometric data.  
$W_0$ consists of Laurent series that represent (via the local 
trivialization of $E$) a holomorphic section of $E$ over 
$\Gamma \setminus \{P_0\}$.  The vacuum $W_0$ is then 
``dressed'' by a Laurent series $\phi(z)$, which is related 
to changing the local trivialization of $E$. 
These geometric data are familiar stuff in the theories of 
algebro-geometric solutions of soliton equations, 
commutative rings of differential operators, etc.  
\cite{bib:Kr78,bib:KN78,bib:KN80,bib:LM97,bib:Mu84,bib:Mu90,
bib:PW89,bib:SW85}.  

This geometric point of view is already enough to tackle 
more general cases.  It is rather straightforward to 
generalize the result for the elliptic analogue of 
the nonlinear Schr\"odinger hierarchy to higher genera, 
though explicit formulas of the $A$-matrices are not 
available therein.   The work of Li and Mulase 
\cite{bib:Mu90,bib:LM97}, too, provides valuable 
material to this issue.

\subsection*{Acknowledgements}

I would like to thank the organizers of the workshop, 
in particular, Nenad Manojlovic and Henning Samtleben, 
for invitation and hospitality. 
This work was partly supported by 
the Grant-in-Aid for Scientific Research (No. 14540172) 
from the Ministry of Education, Culture, 
Sports and Technology.

\end{document}